\documentclass{aa}
\usepackage{epsfig}
\newcommand{\etal}{et al. }
\newcommand{\grs}{GRS 1915+105 }
\begin{document}

\title{Evidence for Synchrotron Bubbles from GRS 1915+105}
\titlerunning{Synchrotron Bubbles from GRS 1915+105}
\authorrunning{Ishwara-Chandra, Yadav \& Pramesh Rao}
\author{C. H. Ishwara-Chandra$^1$, J. S. Yadav$^2$ \and A. Pramesh
Rao$^3$}
\institute{$^1$ Space Astronomy and Instrumentation Division, 
ISRO Satellite Center, Bangalore - 560 017, INDIA \\
$^2$ Tata Institute of Fundamental Research, Homi Bhaba Road,
Mumbai - 400 005, INDIA \\
$^3$ National Center for Radio Astrophysics, Post Bag No. 3,
Ganeshkhind, Pune - 411 007, INDIA  \\
}
\offprints{CHIC; email: ishwar@ncra.tifr.res.in}

\date{}

\abstract{We present GMRT observations of the Galactic microquasar 
\grs at 1.28 GHz for 8 days from 2001 June 18 to July 1. 
We have seen several isolated radio flares of varying magnitudes 
(20 - 50 mJy) and durations (6 - 35  min)  and we model them as 
due to adiabatically expanding synchrotron  emitting  clouds 
(synchrotron bubbles) ejected from the accretion disk. By  applying  
this model, we provide a new method to estimate the electron power-law  
index $p$, hence the spectral index, from single frequency radio  
observations. This method does not require correction for  the optical 
time delay effects  which may be important in the case of optically thick 
radio emission.  Using our  estimated value of $p$ and 
simultaneous multiwavelength data from literature,  we have 
calculated the time of ejection of the synchrotron plasma and 
the time delays at different observed frequencies. 
Our estimates are in good agreement with the observed time delays.  


}

\maketitle

\section{Introduction}
The Galactic X-ray transient source \grs was discovered in 1992 by 
the {\tt GRANAT} satellite (Castro-Tirado \etal 1992). It has now 
been established as a black hole binary with the black hole mass of 
14$\pm 4 M_\odot$ and the companion K-M III star of mass 
1.2$\pm 0.2 M_\odot$ (Greiner \etal 2001). The source was shown 
to possess relativistic outflows during its radio outbursts 
(Mirabel \& Rodr\'iguez 1994). Extensive monitoring of \grs in 
the radio  and X-ray band showed that the overall radio emission is 
correlated  with its X-ray properties (eg: Harmon \etal 1997). 
The radio emission can be broadly classified into three classes; 
(i) the relativistic superluminal radio jets of flux density 
$\sim$ 1 Jy with decay time-scales of several days (Fender \etal 1999),
(ii) the baby jets of 20 $-$ 40 min durations with flux density of 
20 $-$ 200 mJy both in infrared (IR) and radio (Pooley \& Fender 1997; 
Eikenberry \etal 1998) and (iii) the plateau state with persistent 
radio emission of 20 $-$ 100 mJy for extended durations (Muno \etal 2001). 
In the case of superluminal jets, the radio emission has steep spectra 
and are observed at large distances (400 $-$ 5000 AU) from the 
accretion disk (Fender \etal 1999; Dhawan \etal 2000).  The radio 
emission at this distance is believed to be decoupled from the accretion
disk (Yadav 2001; Muno \etal 2001). In contrast, the other two classes
of radio emission has flat spectra and occur close to the accretion
disk (within a few tens of AU). Even though superluminal jets and
baby jets are differentiated by their spectra, decay time scales,
and the distance (from the accretion disk) where the emission occurs, 
there is evidence for ejection of significant amount of relativistic 
material even during the baby jets (Fender \etal 1999). The radio 
emission is believed to be due to synchrotron emission from 
relativistic electrons and the dominant decay mechanism is the 
adiabatic expansion losses (Mirabel \etal 1998, hereinafter M98). 
Feroci et al. (1999) have studied an isolated radio flare 
($\sim$ 60 mJy at 15 GHz) in X-ray and radio and present strong
evidence for mass ejection (QPO disappears during the radio outburst).
Coordinated multi-wavelength studies have indicated strong 
disk-jet connection (M98; Eikenberry \etal 1998; Yadav 2001).

The time of the ejection of the radio emitting synchrotron plasma from the
accretion disk is an important parameter to understand the 
disk$-$jet connection. Another
important parameter is the power-law index ($p$) of the electron energy
distribution in the plasma which is related to the delay time. Even though
simultaneous multi-wavelength radio observations provide $p$, in the case
of optically thick radio emission it is affected by the optical depth time
delay effects and may introduce large error in the estimate of time of
ejection, particularly in the case of periodic emission like IR and 
radio baby jets.

In this {\it Letter} we provide a new method, by applying the adiabatic
expansion model to baby jets to estimate $p$ from single frequency
radio observations.  This method has the unique advantage in the sense
that it  surpasses the optical depth effects.   We apply this method
to isolated baby jets seen in  the radio observations of \grs with the
Giant Meterwave Radio Telescope (GMRT) at 1.28 GHz  and estimate $p$
and the time of ejection of the synchrotron  bubble. Using this $p$
and the multiwavelength data from literature we also calculate the
time delays at different observed frequencies and the time at which the
synchrotron bubble is ejected. 
In Section 2 we present the observational details and basic results.
The adiabatic expansion model is given in Section 3. 
Discussions and conclusions are presented in Section 4.

\section{Observations and Results}  
The radio observations were carried out at 1.28 GHz with a
bandwidth of 16 MHz using the GMRT (Swarup \etal 1991) on 2001 June
18, 22, 23, 27, 28, 29, 30 and July 1.  The flux density scale is
set by observing the primary calibrator 3C286 or 3C48. A phase 
calibrator was observed before and after a 45 min scan on 
GRS 1915+105. The integration time was 32 s. The data recorded from
GMRT has been converted to FITS and was analysed using Astronomical
Image Processing System ({\tt AIPS}). 
The lightcurve produced at 5 min interval is shown in Figure 1
(top panel).  \grs exhibited significant radio emission on all days,
except on June 27. On June 18, the source was quite steady with 
the flux density $\sim$ 10 mJy.
However, on June 22 the source became bright ($\sim$ 50 mJy) and showed
rapid drop in the flux density by a factor of two and slowly recovered to
about the same level.  On June 23, the flux density was steady at about
30 mJy. The source became very weak in radio ($\sim$ 5 mJy) on June 27.
However on 28, the flux density was $\sim$ 5 mJy at about UTC 17 hour
(beginning of observation) 
and reached gradually to 50 mJy at UTC 24 hour. When the observations
resumed on 29, the source was ``caught" at 70 mJy at UTC 16 hour, and the
flux started decaying slowly to the value of 10 mJy at UTC 24 hour. The
source was relatively faint on June 30 and July 1.

We have checked the radio data at shorter binning time of 1 min, which
showed clear short term variations. Minor flares of amplitude upto 50 mJy
have been seen on June 28, 29, 30 and July 1, which are typical of baby
jets (Figure 3 and Table 1). The X-ray hardness ratio (5-12 keV/1.5-5
keV) from RXTE/ASM shown in the bottom panel of Figure 1 suggests that
overall  \grs remained in the low hard state. 

\begin{figure}[t]
\epsfig{figure=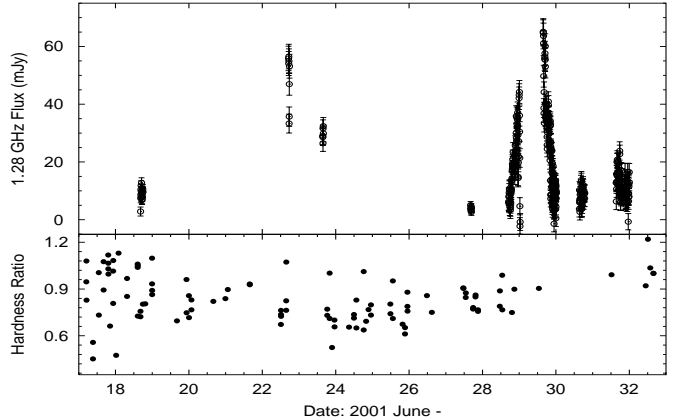,width=5.5cm,height=8.7cm,angle=-90}
\caption{{\it Top panel:} Radio light curve from GMRT at 1.28 GHz,
binned at 5 min interval. The observations on 2001 June 18, 22, 23 and
27 is for $\sim 30 -$ 60 min, while rest of the days are
of $\sim 8 - $10 hours. {\it Bottom panel:} RXTE/ASM hardness ratio 
(5-12 keV/1.5-5 keV) for the same duration.  
}
\label{fig1}
\end{figure}

\section{Adiabatic Expansion Model}

We propose that the minor flares (baby jets) seen in the GMRT 
observations are due to adiabatically expanding synchrotron emitting 
bubbles. Hjellming \& Johnston (1988) present models for explaining 
the persistent radio emission from X-ray binaries by assuming adiabatic 
expansion dominated by lateral motions. For flaring events, they consider 
spherical bubbles of relativistic plasma expanding under its pressure.
We consider this simple adiabatically expanding sphere model 
(three dimensional expansion) for the baby jets seen in the GMRT 
observations to understand their overall radio behavior. We assume 
that the expansion is uniform and the transition time from optically 
thick to thin state is negligible.

Consider a spherical cloud of relativistic electrons with the energy 
distribution $N(E)dE \propto E^{-p}dE,$ emitting synchrotron radiation. 
If this cloud is expanding adiabatically, the observed flux density 
as a function of time will be (van der Laan, 1966; 
Hjellming \& Johnston, 1988)

\begin{equation}
S_{\rm p}/S_{\rm s} = (r_{\rm p}/r_{\rm s})^3
\end{equation}
under optically thick conditions, and
\begin{equation}
S_{\rm e}/S_{\rm p} = (r_{\rm e}/r_{\rm p})^{-2p}
\end{equation}
under optically thin conditions. Here $S_{\rm s}, S_{\rm p}$ and $S_{\rm e}$ are the
observed flux densities at the start, peak and the end of
the flare and $r_{\rm s}, r_{\rm p}$ and $r_{\rm e}$ are the corresponding radii of
the bubble. $p$ is the power-law index of electron energy
distribution.  We define the time sequence of synchrotron bubble as in
Figure 2.  $t_{\rm s}$ is the time taken for the bubble to attain the flux
density $S_{\rm s}$, $t_{\rm p}$ is the time taken since $t_{\rm s}$ for the flux density to
reach its peak value $S_{\rm p}$ (expansion under optically thick conditions)
and $t_{\rm e}$ is the time taken since $t_{\rm p}$ by the bubble to decay to the
flux level of $S_{\rm e}$ (expansion under optically thin conditions). 
Assuming uniform expansion, the profiles will be
\begin{equation}
S_{\rm p} = S_{\rm s}[1 + (t_{\rm p}/t_{\rm s})]^3
\end{equation}
for the rising part of the light curve and
\begin{equation}
S_{\rm e} = S_{\rm p} \{1 + [t_{\rm e}/(t_{\rm p} + t_{\rm s})]\}^{-2p}
\end{equation}
for the decay part. An interesting case is when the synchrotron bubble
is isolated, where the rise, peak and decay of a single bubble is
observed. The adiabatic equations (1) and (2) can be solved to obtain $p$

\begin{equation}
p = {1 \over 2}{{\rm log}(S_{\rm p}/S_{\rm e}) \over {\rm log}\{1 + (t_{\rm e}/t_{\rm p})[ 1 -
(S_{\rm p}/S_{\rm s})^{-1/3}]\}}
\end{equation}
Thus by measuring the rise and decay time from isolated flares, the
electron power-law index $p$ can be obtained.  For estimating $p$, we
have considered only those flares where the rise and decay of single
flare is clearly visible well above the noise level or steady emission.
We fit the above equations (using the least square method) to such 
isolated flares seen in our data on 2001 June 28, 29, 30 and July 1. 
The best fit model for two such flares are given in Figure 3. 
The estimated parameters are given in Table 1.  
Column 1 gives the date of observation, Column 2 gives the 
estimated time (UT) of birth of the synchrotron bubble 
(see Section 4 for details).  The duration and peak flux density 
of the flares are given in Columns 3 and 4. The electron power-law 
index ($p$) and the spectral index ($\alpha = (p - 1)/2$) are 
given in Columns 5 and 6.

\begin{figure}[t]
\vbox{
\epsfig{figure=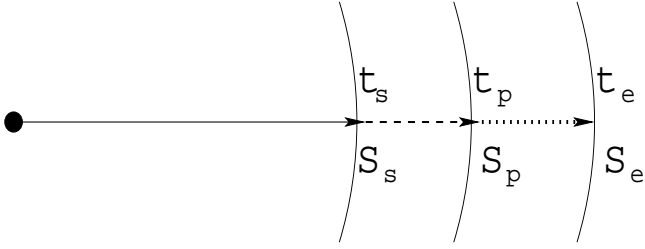,height=8.5cm,width=3.2cm,angle=-90}
}
\caption{Illustration of the time markers during the adiabatic
expansion. The point on the left marks the ejection time.
}
\end{figure}

\begin{figure}[t]
\vbox{
\epsfig{figure=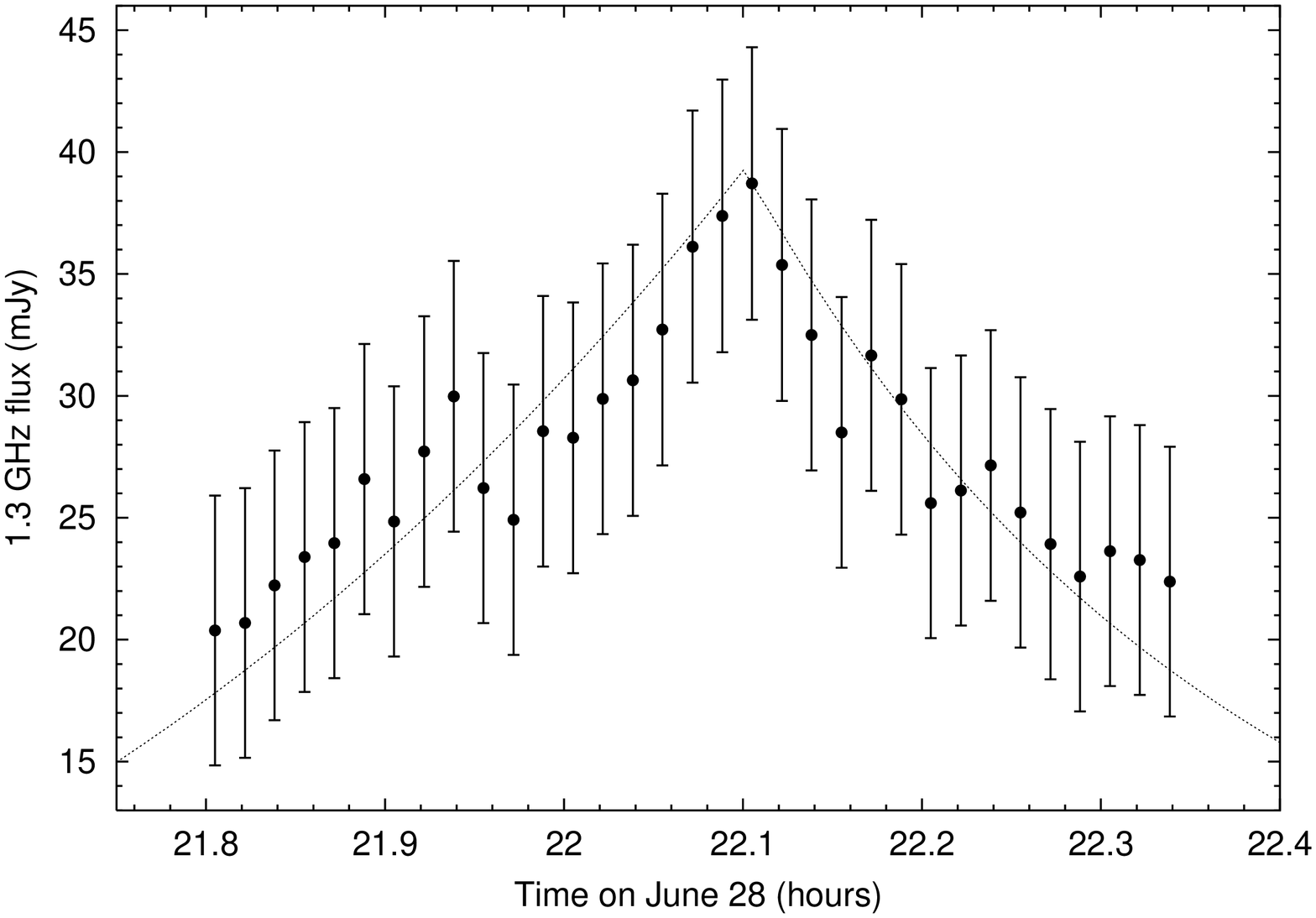,width=5.4cm,height=8.7cm,angle=-90}
}
\vspace{0.25cm}
\vbox{
\epsfig{figure=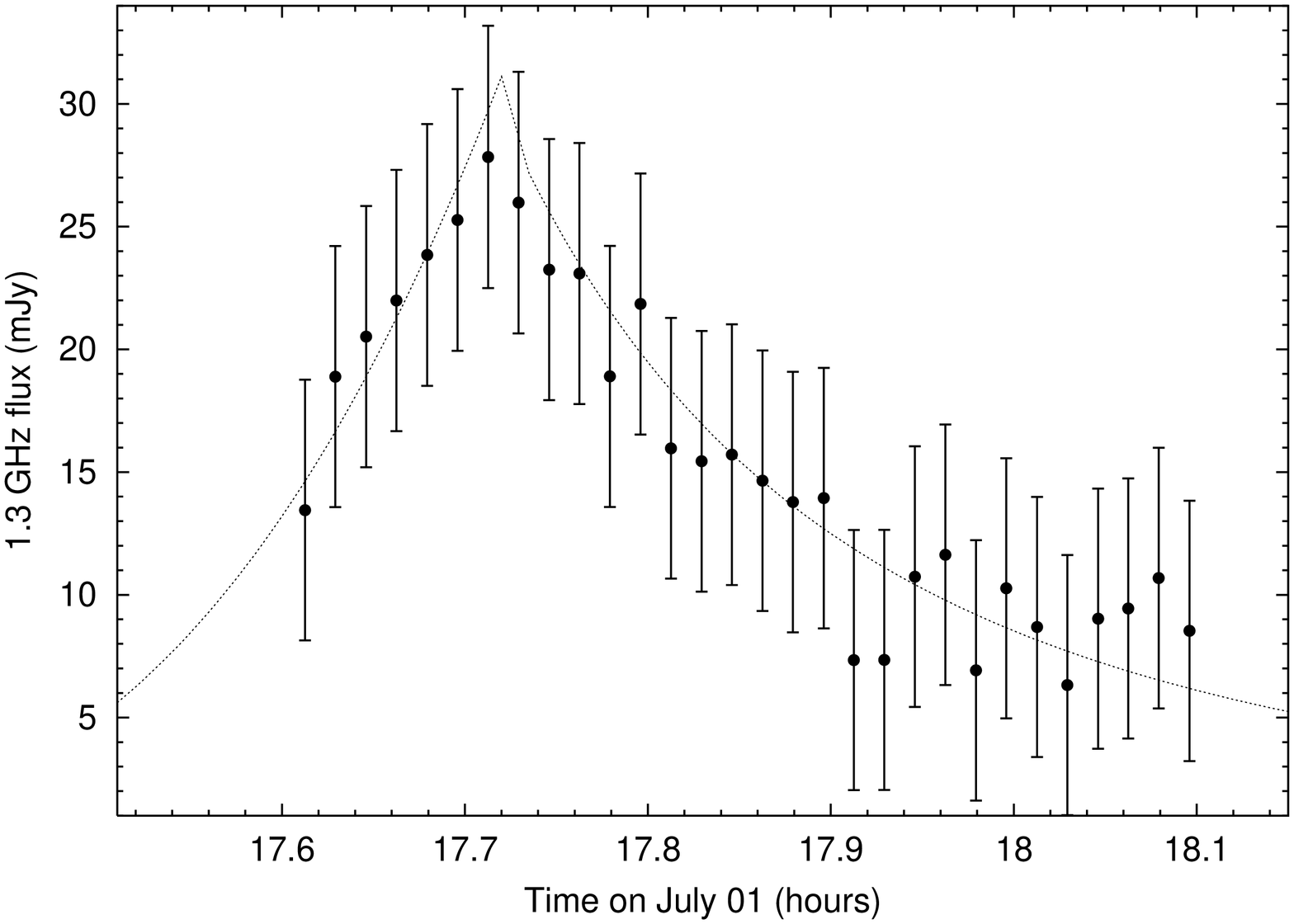,width=5.4cm,height=8.6cm,angle=-90}
}
\caption{Examples of isolated flares from GMRT observations. 
{\it Top panel:} The isolated flare seen on 2001 June 28.  
{\it Bottom panel:} One of
the four minor isolated flares seen on July 1.
The data is binned at 1 min interval, the time is in UT. The
dotted line is the fit to the equations (3) and (4) for the rising part
and for the optically thin decay respectively.
}
\label{fig3}
\end{figure}

\section{Discussion and Conclusion}

We have presented the GMRT observations of the micro-quasar \grs at 1.28
GHz. Several isolated minor flares which are typical of baby jets have
been found. We have suggested that these flares are due to adiabatically
expanding bubbles of synchrotron emitting plasma. By applying adiabatic
expansion model we have provided a new method to estimate the electron
power-law index $p$ from single frequency radio observations, which
does not need correction for the optical depth time delay effects.
The simultaneous multi-wavelength radio observations of \grs reported by
Fender \etal (2002) reveal short time large variations in the spectral
index from optically thick ($\alpha \sim -0.2$)\footnote{We define 
the radio spectral index $\alpha$ as $S_\nu \propto \nu^{-\alpha}$}
to thin values ($\alpha \sim 0.7$) which corresponds to $p$ of 2.4 in the
optically thin regime.  They observed the fast oscillations of about
half an hour period on MJD 51751.  The power-law index $p$ obtained from
our observations (Table 1) are in agreement with the measured values of Fender
\etal (2002) in the optically thin region. The simplest explanation
to this is that at the peak of the oscillation, the contribution from
the optically thick radio emission is large and hence the delay time
correction becomes important. However, at low flux level the contribution
from optically thick emission is reduced and the delay time correction
is less important. This explanation for variations in $\alpha$ 
reported by Fender et al (2002) suggests that intrinsically $p$ 
could be constant which is consistent with an adiabatically expanding
synchrotron source with no additional acceleration. It is difficult
to visualise change in $p$ over such a short duration of $\sim$ 15 min.

We can calculate the time delays between emission at different wavelengths
using this $p$.  The time at which the emission at an wavelength $\lambda$
reaches its maximum is given by (van der Laan, 1966) 
\begin{equation}
t_{\rm m,\lambda} \propto \lambda^{{p + 4 \over 4p + 6}} 
\end{equation}

We calculate the proportionality constant of equation 6 using our
estimate of $p$ and the  time delay of 0.25 h between 8.3 GHz and 5
GHz radio emission observed  in \grs by M98. The value of $t_{\rm 6cm}$
estimated by us for  May 15, 1997 observations of M98 is $\sim$1.2 h
(M98 gives 0.9 h assuming $p$ = 0) and the ejection of plasma occurred
at 14.95 $-$ 1.2 = 13.75 h.  Inspection of the X-ray data at this time
(Figure 1 of M98) shows that it still falls in the gap in the X-ray data.
We have calculated the time of formation of the synchrotron bubble for
the flares observed by us  which are given in Table 1.

\begin{table}[t]
\caption{Estimated parameters for the synchrotron bubbles from 
GMRT Observations at 1.28 GHz.}
\begin{tabular}{cccccc}
Obsn    & Time of  & Dur   & Peak &$p$            & $\alpha$  \\
date    & ejection & (min)    & Flux &          &                \\
&          &        &      &         &                 \\
06/28   &19.8126   &$\sim$ 30& 38.8 & 2.17$\pm$0.17 & 0.58$\pm$0.09  \\
06/29   &14.0653   &$\sim$ 8 & 51.3 & 1.25$\pm$0.12 & 0.13$\pm$0.06  \\
06/30   &18.0620   &$\sim$ 30& 30.1 & 1.64$\pm$0.08 & 0.32$\pm$0.04  \\
07/01   &15.5077   &$\sim$ 35& 27.9 & 1.36$\pm$0.07 & 0.18$\pm$0.04  \\
07/01   &20.3278   &$\sim$ 12& 19.4 & 1.51$\pm$0.13 & 0.25$\pm$0.06  \\
07/01   &21.5200   &$\sim$  9& 20.5 & 1.12$\pm$0.11 & 0.06$\pm$0.05  \\
07/01   &21.6818   &$\sim$  6& 32.8 & 2.01$\pm$0.32 & 0.50$\pm$0.16
\end{tabular}
\vspace{-3mm}
\end{table}

We give in Table 2 the observed time delays between different
frequencies from simultaneous multiwavelength observations of
\grs reported in the literature. We have estimated  the time delays 
between these  frequencies using the weighted mean value of 
$p$ = 1.46 from our observations which are also given in Table 2.
For observations on MJD 50675, the calculated time delay of 37 s for the
IR emission is consistent with the observation that the spikes in X-ray
coincide with the beginning of IR flares (Eikenberry \etal 1998).  The
time delay of 200-400 s reported by them between the X-ray and IR peaks
is probably due to the time required for the evolution of the IR flare
as well as for the readjustment of the accretion disk.  Our calculated
time delay of 804 s is in good agreement with the measured time delay
between 15 GHz and 8.3 GHz observations reported by M98 (MJD 50583). 
In a detailed study of X-ray - radio observations in \grs, 
Klein-Wolt et al. (2002) measure the time delay of 40 $-$ 45 min 
from the beginning of state C to the peak radio emission at 15 GHz,
which agrees with our estimate of 43 min. The observations of 
M98 (MJD 50700) has measured time delay of $\sim$ 960 s between 
IR and 8.3 GHz, while our calculated value is 3351 s (0.93h),
the difference is about 40 min. Inspection of Figure 2 of M98 show that
a previous IR peak is separated by about the same amount of time. This
suggests that the 8.3 GHz flare most likely corresponds to the IR flare
$\sim$ 0.9 h prior to the 8.3 GHz peak. For observation on MJD 50705,
the discrepancy of 547 s between the measured and the calculated time
delay between IR and radio emission could be due to larger uncertainty
in the measured time delay because of the presence of two IR peaks. 
From the radio observations at 4.8 and 8.6 GHz on MJD 51751, 
Fender \etal (2002)
find a delay time of $\sim$ 600 s based on a broad peak ($\sim$ 300 s)
in the cross correlation function while our estimate of time delay is
1030 s. Although the exact answer to this discrepancy is not clear,
it may be due to uncertainty in the time delay measurements. 

\begin{table}[t]
\caption{Calculated time delays at different frequencies for the 
multiwavelength data from literature.}
\begin{tabular}{c c c c c c}
MJD   &$\nu_1$  & $\nu_2$ & $\Delta t$ & $\Delta t$ & Ref \\
      & (Hz)    & (Hz)    & (obs) & (calc) &     \\
      &         &         &        &        &     \\
50675 & X-ray              & 1.4$\times10^{14}$& 200$-$400 & 37 & E98 \\
50583 & 1.5$\times10^{10}$ & 8.3$\times10^9$ &$\sim$750 & 804 & M98 \\
50698 & X-ray              & 10$^{15}$       & 2400$-$2700$^*$ & 2584 & K02\\
50700 & 1.4$\times10^{14}$ & 8.3$\times10^9$ &$\sim$960 & 3351& M98 \\
50705 & 1.4$\times10^{14}$ & 1.5$\times10^{10}$&$\sim$2000 & 2547 & F98\\
51751 & 8.6$\times10^9$    & 4.8$\times10^9$   & $\sim$600 & 1030 & F02 \\
\multicolumn{6}{l}{$^*$When calculated from the beginning of State C}
\end{tabular}
\vspace{-2mm}
\end{table}

It may be noted here that the time delays estimated using our simple 
model of synchrotron bubble are in good agreement with many of the
observed values between different frequencies available 
in the literature. However, naive application of this model
predicts much higher flux densities than observed at higher frequencies.
This discrepancy may be the subject of future studies. \\[0.5mm]

\noindent{\it Acknowledgments:} We thank the referee, Dr. G. G. Pooley
for meticulously reading the paper and making important comments.
We also thank several of our colleagues, especially Dr. A. K. Jain and 
Prof. A. R. Rao for stimulating discussions.  We thank the staff of the 
GMRT that made these observations possible.
GMRT is run by the National Center of Astrophysics of the Tata Institute
of Fundamental Research. This research has made use of data 
obtained from the High Energy Astrophysics Science Archive Center
(HEASARC), provided by NASA's Goddard Space Flight Center and also of
NASA's Astrophysics Data System Abstract Service.

\end{document}